\documentclass{article}
\usepackage{spconf,amsmath, amssymb, graphicx,hyperref}
\usepackage{algorithm}       
\usepackage{algpseudocode}    % algorithmicx + \Require/\Ensure/\Comment
\usepackage{booktabs}
\usepackage{multirow}

\makeatletter
\renewcommand{\fnum@algorithm}{\normalsize\bfseries Algorithm~\thealgorithm}
\makeatother

\algrenewcommand\algorithmicrequire{\textbf{Input:}}
\algrenewcommand\algorithmicensure{\textbf{Output:}}
\algrenewcommand\algorithmiccomment[1]{\hfill$\triangleright$~#1}

\vspace{-10cm}
\title{A mixed precision FFT with applications in MRI}
\name{%
Nikhil Deveshwar$^{1,2}$,
Abhejit Rajagopal$^{3}$,
Peder E.Z. Larson$^{1,2}$%
}

\address{%
$^{1}$UC Berkeley-UCSF Graduate Program in Bioengineering\\
$^{2}$Department of Radiology and Biomedical Imaging, University of California San Francisco\\
$^{3}$Department of Urology, University of California San Francisco and Allen Institute\\
}

\begin{document}
%\ninept
%
\maketitle
\begin{abstract}
A mixed precision Fast Fourier transform (FFT) implementation is presented. The procedure uses per-block microscaling (MX), a global power-of-two prescale, and prequantized low-bit twiddles. We evaluate forward and round-trip FFT fidelity on two public MRI datasets and compare the effect of various low precision formats, image sizes, and MX block sizes on image quality. Results show that mantissa precision is the primary limiter under MX scaling while ablations suggest weak dependence on image size but a clear block-size trade-off with larger block sizes resulting in better numerical performance.
\end{abstract}
\begin{keywords}
FFT, mixed precision, microscaling, 8-bit floating point, MRI
\end{keywords}

\vspace{-3mm}
\section{Introduction}
\label{sec:intro}

%\section*{Introduction}
The Fast Fourier transform (FFT) is ubiquitous in a wide range of signal and image processing pipelines, including MR imaging. To reduce memory traffic and energy consumption on edge and embedded platforms such as low cost and portable MRI scanners \cite{Wald2019, Liu2021}, there is a growing interest in low-precision and mixed-precision arithmetic.

8-bit floating point (FP8) standards have been recently deployed via NVIDIA’s Transformer Engine, which applies per-tensor/channel scaling with an amax history for stable FP8 training \cite{micikevicius_fp8_2022, nvidia_te_docs_2025}. The Open Compute Project (OCP) has also standardized microscaling (MX) formats and block-shared-exponent encodings such as MXFP8/6/4 \cite{ocp_ofp8_2023, ocp_mx_2023, rouhani_mx_formats_2023}. On the other hand, lower precision FFTs have historically used block floating-point (BFP) implementations on DSP/FPGA platforms \cite{ti_bfpfft_spra948_2002} while modern GPU stacks use FP16/BF16 FFT implementations \cite{nvidia_cufft_13_half_2025} with more recent work focusing on further accelerated FP16 FFT on tensor cores 
\cite{li_tcfft_2021}. 

However, low-precision naive quantization in complex-valued FFTs can induce overflow/underflow errors and accuracy loss. Here, we introduce MX-style scaling and FP8 quantization to the FFT with a focus on numerical accuracy. Our implementation adapts MX-style per-block scaling (common in FP8 GEMM kernels \cite{rouhani_mx_formats_2023}) to the complex multiplication  butterfly operations with pre-quantized FP8 twiddles, FP32 accumulation (common in GPU mixed precision kernels \cite{nvidia_te_docs_2025}), and a single global power-of-two prescale to set the initial range of the input. Previous studies \cite{Mirsalari2024} have broadly characterized FP8 performance by benchmarking various DSP/ML kernels, however this work is limited to static scaling or bias shifts nor is it tied to a specific application. Our approach is validated on forward and round-trip workloads on MRI data, a classic application where the FFT is used to convert acquired frequency domain data (k-space) to image space. This bridges classic BFP-FFT ideas with modern FP8/MX implementations, but applied to an imaging workload where mantissa vs range trade-offs can behave differently than in GEMM kernels.

% Specifically, our contributions are as follows:
% \begin{itemize}
%     \item \textbf{MX-FP8 butterfly kernel} A radix-2 iterative FFT where the complex multiply in each butterfly is executed in MX-FP8 (e4m3/e5m2) with pre-quantized FP8 twiddles
%     \item \textbf{Stability scaling} A single power-of-two global prescale and stage-wise MX re-encoding to mitigate overflow/underflow.
%     \item \textbf{MRI Validation} Forward and roundtrip FFT assessments on two different MRI datasets (anatomy \& sequence diversity)
%     \item \textbf {Ablation} Effects of image size and MX block size $k$ on forward FFT.
% \end{itemize}

\begin{figure*}[h]
  \centering
  \includegraphics[scale=0.25]{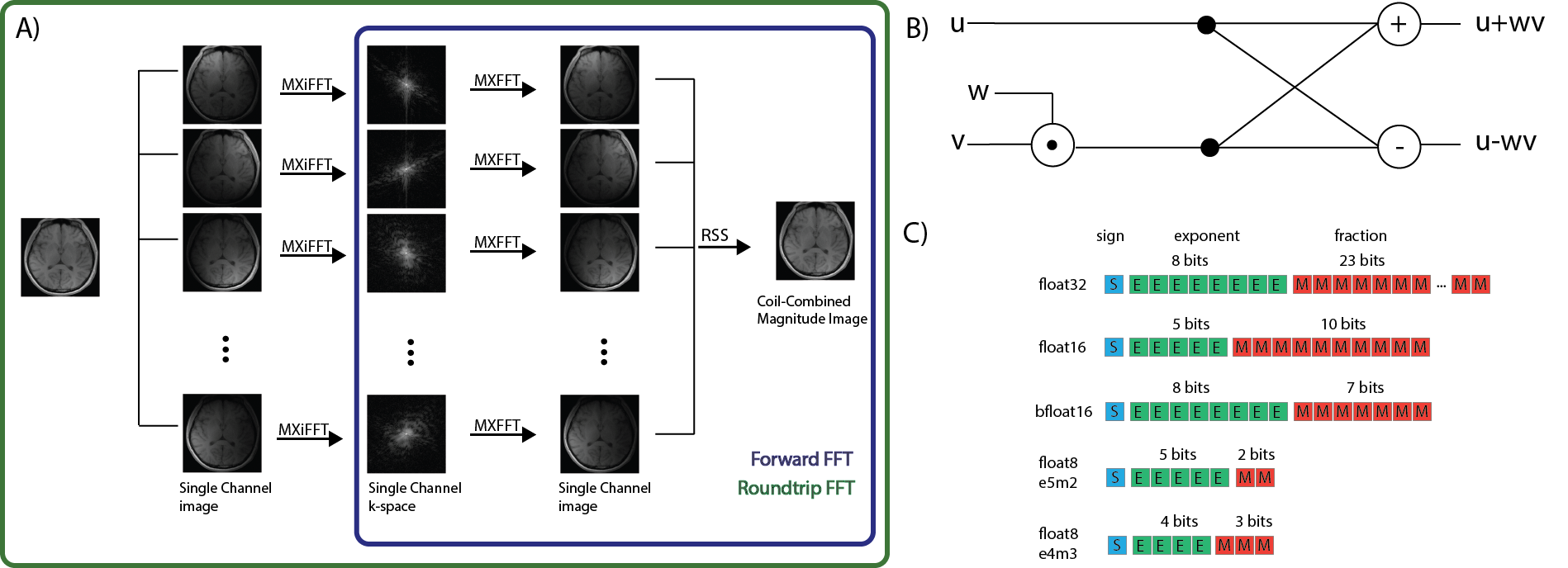}
  \caption{A: Forward/Round-trip FFT for MRI, B: Butterfly Operations, C: Floating point formats; E=exponent, M=mantissa bits}
  \label{fig:overview}
\end{figure*}

\vspace{-5mm}

\section{Methods}
\label{sec:methods}

%\section*{Methods}

\begin{figure*}[t]
  \centering
  \includegraphics[scale=0.45]{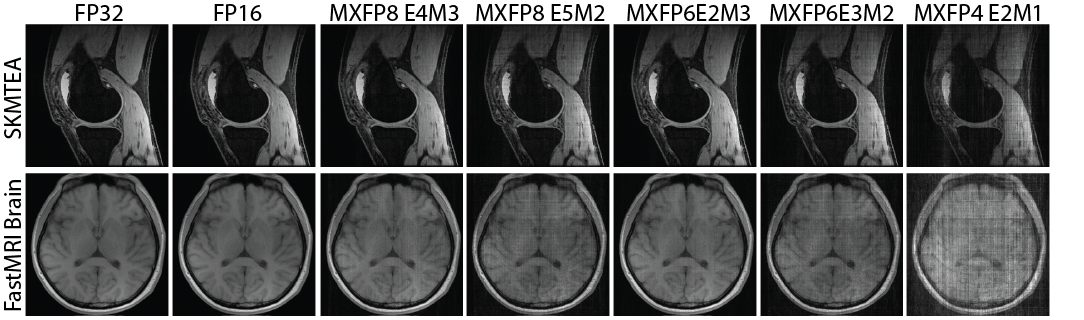}
  \caption{Forward MXFFT in low precision MX variants}
  \label{fig:forward_fft}
\end{figure*}
% We first describe a global scaling procedure and  the miciroscaling (MX) implementation for the butterfly complex multiplication. The evaluation methods and datasets are then discussed.
\subsection{Global Prescale}
To avoid overflow/underflow errors in the butterfly operations, we first apply a single power-of-two global prescale before computing the FFT to keep peaks of the input near a desired target and lift small-magnitude tails above a user-defined floor. Once the global peak magnitude and $\tau$-percentile of the nonzero entries of the input are set, two candidate exponents are computed to place the peak at the desired target ($k_1$) and to raise the tail to $\tau_{min}$ ($k_2$). The applied gain uses the stricter requirement with user-supplied clipping if desired. The input is then scaled with applied gain ($k$) with the \texttt{ldexp} function. Alg \ref{alg:global_prescale} outlines the procedure.

%----------NON IEEE FRIENDLY------------------
% \begin{algorithm}[t]
% \caption{Global power-of-two prescale}
% \begin{algorithmic}[1]
% \Require $x$ (complex), format flag $f\in\{\mathrm{e4m3},\mathrm{e5m2}\}$, target peak $\text{target}$, tail percentile $\tau$, optional tail floor $\tau_{\min}$, bounds $(k_{\min},k_{\max})$
% \Ensure $x_s$ (scaled input)
% \State $m \gets |x|$;\; $a_{\max} \gets \max(m)$;\; $nz \gets \{m_i \mid m_i > 0\}$;\; $p_\tau \gets \mathrm{Percentile}(nz,\tau)$ 
% % \If{$\neg$finite$(a_{\max})$ \textbf{or} $a_{\max}=0$}
% %   \State \Return $x, 0$
% % \EndIf
% % \If{$\tau_{\min}$ not provided}
% %   \State $\tau_{\min} \gets \begin{cases}
% %     2^{-10}, & f=\mathrm{e4m3}\\
% %     2^{-12}, & f=\mathrm{e5m2}
% %   \end{cases}$
% % \EndIf
% \State $\epsilon \gets 10^{-30}$
% \State $k_1 \gets \operatorname{round/}\!\log_2\!\big(\text{target}/\max(a_{\max},\epsilon)\big)$
% \State $k_2 \gets \left\lceil \log_2\!\big(\tau_{\min}/\max(p_\tau,\epsilon)\big) \right\rceil$
% \State $k \gets \operatorname{clip}\!\big(\max(k_1,k_2),\,k_{\min},\,k_{\max}\big)$
% \State $x_s \gets \operatorname{ldexp}(x, k)$ %\Comment{exact $x \gets 2^k x$}
% \State \Return $x_s$
% \label{alg:global_prescale}
% \end{algorithmic}
% \end{algorithm}

\begin{algorithm}[h]
\caption{Global power-of-two prescale}
\label{alg:global_prescale}
\small
\begin{algorithmic}[1]
\Require $x\in\mathbb{C}^{N \times N}$, target peak $\mathrm{target}$, tail percentile $\tau$, floor $\tau_{\min}$, bounds $(k_{\min},k_{\max})$
\Ensure $x_s \in \mathbb{C}^{N \times N}$ 
\State $m \gets |x|$; $a_{\max}\gets \max(m)$; $m_{nz}\gets \{m_i\,:\,m_i>0\}$; $p_\tau\gets \mathrm{Percentile}(m_{nz},\tau)$ \Comment{get global peak $a_{\max}$ and $\tau$-percentile of nonzeros }
\State $\epsilon\gets 10^{-30}$ \Comment{Avoid $\log_2(0)$ if peaks or tails are zero}
\State $k_1\gets \mathrm{round}\!\left(\log_2\!\frac{\mathrm{target}}{\max(a_{\max},\epsilon)}\right)$ \Comment{Exponent to place the peak at the target} 
\State $k_2\gets \left\lceil \log_2\!\frac{\tau_{\min}} {\max(p_\tau,\epsilon)} \right\rceil$ \Comment{Minimum exponent to lift tail to the floor (prevent underflow)}
\State $k\gets \mathrm{clip}\!\big(\max(k_1,k_2),\,k_{\min},\,k_{\max}\big)$ \Comment{Use stricter requirement and clamp if specified}
\State $x_s\gets \mathrm{ldexp}(x, k)$ \Comment{Apply scale $x_s=2^{k}x$}
\State \Return $x_s$
\end{algorithmic}
\end{algorithm}

%----- step by step -------
% require...
% measure current dynamic range, get amax of input magnitude, and tau percentile of nonzeros
% if tau_min no provided, choose one based on expected mxfp format for the tail floor
% set eps tiny to avoid log(0)
% k1: get pow2 that moves current peak to target pkea
% k2: minimum pow2 to raise tau_percentile to tau_min (don't underflow small but meaningful values)
% clip stricter requiremnt (max(k1,k2))
% apply the scaling as an exact pow2 multiply (no rounding error in binary FP), and return prescaled signal/image
% choose 0.5 target peak so biggest sample ends up at 50% of full scale (this is to lead headroom if there is growth in butterflies and mitigates overflow risk)

\subsection{MX-scaled Complex Butterfly Operations}
Next, we implement the twiddle complex multiply inside each Cooley–Tukey butterfly (Fig \ref{fig:overview}B) using MX block-floating FP8 format (Fig \ref{fig:overview}C) on small complex blocks ($B{=}32$ corresponding to 32 real elements). The approach starts with MX-encoding the operand blocks ($w$ and $v$). The mantissas are then extracted to perform the complex multiplication in mantissa "space". Finally, the product is repackaged as a fresh MX block before the butterfly addition and subtraction which are performed in a higher precision (FP32). If the largest absolute mantissa value produced during the complex multiply is greater than the format mantissa, the MX-product block is re-normalized.

The procedure uses helper functions \texttt{EncodeBlockMX}, \texttt{MantissasBlock}, \texttt{EncodeFromMantBlock}, and \\ \texttt{DecodeBlockMX}, which call the \texttt{gfloat}~\cite{gfloat_docs_v05, gfloat_github} API for MX block quantization. \texttt{EncodeBlockMX} selects a shared block scale $(s_w, s_v, \hat{s})$ and returns FP8 codes $(c_w, c_v, c_p)$. \\ \texttt{MantissasBlock} returns de-scaled mantissas for the real and imaginary components $(x_r, y_r)$ and $(x_i, y_i)$. \\ \texttt{EncodeFromMantBlock} repacks mantissas with a new scale. \texttt{DecodeBlockMX} reconstructs the resultant complex values from the mantissa operation $(wv)$. Alg \ref{alg:butterfly_mx} describes the procedure. As a positive control for low-precision accuracy, we implement a 16-bit FFT with quantized inputs and outputs and butterfly complex multiplications computed in standard FP16 arithmetic with accumulation in FP32.
\vspace{-5mm}

\begin{algorithm}[h]
\caption{MX-scaled Complex Butterfly}
\label{alg:butterfly_mx}
\small
\begin{algorithmic}[1]
\Require $u,v,w\in\mathbb{C}^{B}$ \Comment{inputs, twiddles, block size}
\Ensure $(y_0, y_1)$ \Comment{butterfly sum and difference}
\State $(c_w, s_w, n)\gets \Call{EncodeBlockMX}{w}$ \Comment{FP8/MX packed}
\State $(c_v, s_v, n)\gets \Call{EncodeBlockMX}{v}$ \Comment{FP8/MX packed}
\State $(x_r,x_i)\gets \Call{MantissasBlock}{c_w, s_w, n}$ \Comment{Depack mantissa}
\State $(y_r,y_i)\gets \Call{MantissasBlock}{c_v, s_v, n}$ \Comment{Depack mantissa}
\State $p_r\gets x_r\odot y_r - x_i\odot y_i$;\quad $p_i\gets x_r\odot y_i + x_i\odot y_r$ \Comment{Higher precision MAC in mantissa space}
\State $s_{\mathrm{out}}\gets s_w s_v$;\quad $a_{\max}\gets \max(\|p_r\|_\infty,\|p_i\|_\infty)$
\If{$a_{\max}>M_{\max}$} \Comment{renormalize mantissas if max block value greater than FP8 mantissa-domain limit}
  \State $k\gets \left\lceil \log_2\!\left(\frac{a_{\max}}{M_{\max}}\right)\right\rceil$
  \State $(p_r,p_i)\gets (p_r,p_i)/2^k$;\quad $s_{\mathrm{out}}\gets s_{\mathrm{out}}\,2^k$
\EndIf
\State $(c_p,\hat{s})\gets \Call{EncodeFromMantBlock}{p_r,p_i,s_{\mathrm{out}}}$ \Comment{requantize mantissa as FP8 with shared scale}
\State $wv\gets \Call{DecodeBlockMX}{c_p,\hat{s},n}$ \Comment{decode product block}
\State $y_0\gets u+wv$;\quad $y_1\gets u-wv$ \Comment{Higher precision MAC}
\State \Return $(y_0,y_1)$
\end{algorithmic}
\end{algorithm}

%\vspace{-6mm}

\subsection{Evaluation}
We evaluate the numerical fidelity of the proposed mixed-precision FFT using an MRI-style workload using fully sampled complex-valued k-space data from two public datasets: fastMRI \cite{https://doi.org/10.48550/arxiv.1811.08839} (brain, 2D fast spin-echo) and SKM-TEA \cite{https://doi.org/10.48550/arxiv.2203.06823} (knee, 3D gradient recalled echo). For a typical MRI acquisition, k-space data are acquired from multiple receive array coils and individually converted to image space via FFT. A root-sum-square operation is used to get the final magnitude image. The forward and round trip FFT per coil can be expressed as
\vspace{-1cm}
\begin{equation}
\begin{split}
x_{\text{RSS}} &= \left( \sum_{c=1}^{C} |T_c|^2 \right)^{1/2}, \text{where} \\
T_c &= 
\begin{cases} 
\mathcal{F}\{X_c\} & \text{(forward)} \\
\frac{1}{N^{2}}\mathcal{F}^{-1}\{\mathcal{F}\{x_c\}\} & \text{(round-trip)}
\end{cases}
\end{split}
\end{equation}

where $X_c \in \mathbb{C}^{N \times N}$ is a 2-D single coil element in the frequency domain,  $x_c \in \mathbb{C}^{N \times N}$ is a 2-D coil element in the image domain and $\mathcal{F}$ and $\mathcal{F}^{-1}$ are our proposed mixed-precision forward and inverse FFT implementation respectively. We compare our implementation against the built-in numpy FP32 FFT reference and our FP16 FFT positive control. Standard computational imaging/computer vision metrics peak-signal-to-noise ratio (PSNR), structural similarity index (SSIM \cite{wang2004image}) and normalized mean squared error (NMSE) were evaluated across 10 RSS images from both datasets. Figure \ref{fig:overview}A shows the overall experimental setup.

\vspace{-.5cm}

\section{Results}
\label{sec:results}

%\section{Results}

\begin{figure}[ht]
  \centering
  \includegraphics[scale=0.46]{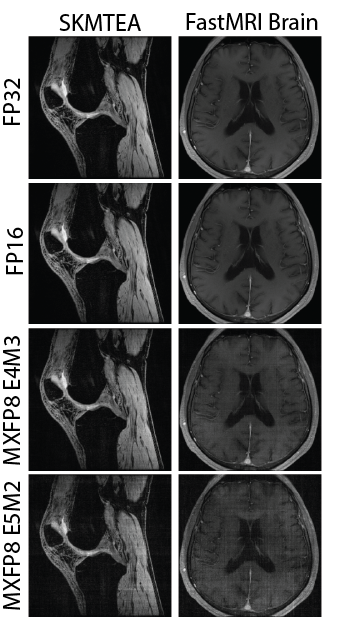}
  \caption{Round-trip MXFFT for both FP8 variants for both datasets}
  \label{fig:roundtrip_fft}
\end{figure}

Figures \ref{fig:forward_fft} and \ref{fig:roundtrip_fft} show forward MX-scaled FFT (MXFFT) and round-trip MXFFT performance on image quality across several low-precision formats. On both datasets, MXFP8-E4M3 delivers the best FP8 performance quantitatively, followed closely by MXFP6-E2M3 for the forward FFT workload. Conversely, formats with only two mantissa bits (E5M2/E3M2) suffer from degraded image quality quantitatively and on visual assessment for both experiments.

\begin{figure}[h]
  \centering
  \includegraphics[width=\columnwidth]{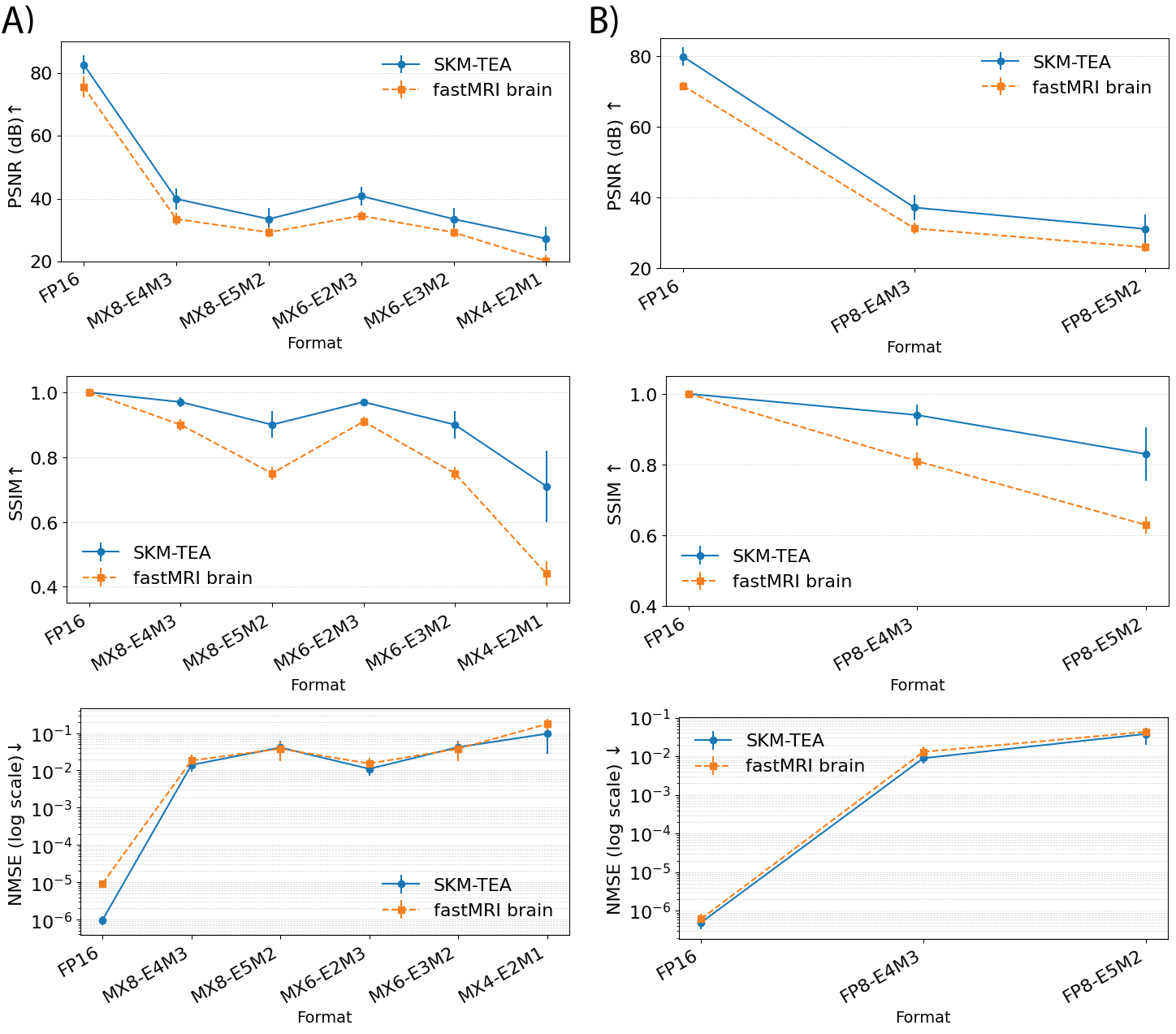}
  \caption{A) Forward FP16 and MXFFT performance across formats. B) Round-trip FP16 and MXFFT quality across formats. Error bars show mean±std over 10 images. MXFFT suffers from higher quantization noise compared to FP16.}
  \label{fig:forward_rt_plots}
\end{figure}

Figure \ref{fig:forward_rt_plots} summarizes forward and round-trip FFT performance across the entire dataset. MXFP8-E4M3 outperforms MXFP8-E5M2 on both fastMRI and SKM-TEA datasets quantitatively. Additionally SKM-TEA consistently shows higher PSNR/SSIM and lower NMSE than fastMRI. Across both datasets and experiments, FP16 FFT serves as an upperbound under low precision with MXFP8-E4M3 and MXFP6-E2M3 closely tracking it for SSIM measurements.

From Table \ref{tab:image_size} the MXFP8-E4M3 format shows slight increase in PSNR as the image size increases from $64 \times 64$ to $256 \times 256$, while SSIM and NMSE remain the same. For MXFP8-E5M2, PSNR/SSIM are roughly flat and NMSE degrades slightly. The increase in image size only adds four like-for-like quantization steps which is probably too small to materially change PSNR/SSIM/NMSE measurements.

Table \ref{tab:block_size} shows the effect of the MX block size $B$. The results suggest block size $B{=}2$ underperforms block size $B{=}\{8, 32\}$. Larger $B$ size can reduce scale/conversion overhead and yields more stable statistics, improving PSNR/NMSE with diminishing returns beyond $B{=}8$ for MXFP8-E5M2 on SKM-TEA. All results use identical prescale target, twiddle quantization, and accumulate precision.

\begin{table}[t]
\centering
\caption{Effect of image size on forward MXFFT performance on SKMTEA dataset}
\label{tab:image_size}
\scriptsize
\setlength{\tabcolsep}{3pt}
\renewcommand{\arraystretch}{0.95}

\textbf{MXFP8 E4M3}\\%[-2pt]
\resizebox{\columnwidth}{!}{%
\begin{tabular}{c c c c}
\toprule
\textbf{Size} & \textbf{PSNR($\uparrow$)} & \textbf{SSIM($\uparrow$)} & \textbf{NMSE($\downarrow$)} \\
\midrule
$64 \times 64$  & $33.7\pm3.35$ & $0.960\pm0.010$ & $5.27\times10^{-3}\pm2.63\times10^{-3}$ \\
$128 \times 128$ & $38.9\pm3.54$ & $0.970\pm0.015$ & $5.31\times10^{-3}\pm1.85\times10^{-3}$ \\
$256 \times 256$ & $40.0\pm3.36$ & $0.970\pm0.014$ & $4.56\times10^{-3}\pm6.94\times10^{-4}$ \\
\bottomrule
\end{tabular}%
}

\vspace{4pt}
\textbf{MXFP8 E5M2}\\%[-2pt]
\resizebox{\columnwidth}{!}{%
\begin{tabular}{c c c c}
\toprule
\textbf{Size} & \textbf{PSNR($\uparrow$)} & \textbf{SSIM($\uparrow$)} & \textbf{NMSE($\downarrow$)} \\
\midrule
$64 \times 64$  & $33.9\pm3.38$ & $0.923\pm0.031$ & $1.43\times10^{-2}\pm2.10\times10^{-3}$ \\
$128 \times 128$ & $33.6\pm3.98$ & $0.900\pm0.046$ & $1.80\times10^{-2}\pm4.33\times10^{-3}$ \\
$256 \times 256$ & $33.58\pm3.46$ & $0.900\pm0.041$ & $2.12\times10^{-2}\pm8.17\times10^{-3}$ \\
\bottomrule
\end{tabular}%
}
\end{table}

\begin{table}[t]
\centering
\caption{Effect of block size on forward MXFFT performance on 128x128 SKMTEA dataset}
\label{tab:block_size}
\scriptsize
\setlength{\tabcolsep}{3pt}
\renewcommand{\arraystretch}{0.95}

\textbf{MXFP8 E4M3}\\%[-2pt]
\resizebox{\columnwidth}{!}{%
\begin{tabular}{c c c c}
\toprule
\textbf{Block $B$} & \textbf{PSNR($\uparrow$)} & \textbf{SSIM($\uparrow$)} & \textbf{NMSE($\downarrow$)} \\
\midrule
2  & $35.9\pm4.15$ & $0.950\pm0.022$ & $1.08\times10^{-2}\pm4.58\times10^{-3}$ \\
8  & $38.6\pm3.98$ & $0.970\pm0.015$ & $5.75\times10^{-3}\pm2.18\times10^{-3}$ \\
32 & $40.0\pm3.36$ & $0.970\pm0.014$ & $4.56\times10^{-3}\pm6.94\times10^{-4}$ \\
\bottomrule
\end{tabular}%
}

\vspace{4pt}
\textbf{MXFP8 E5M2}\\%[-2pt]
\resizebox{\columnwidth}{!}{%
\begin{tabular}{c c c c}
\toprule
\textbf{Block $B$} & \textbf{PSNR($\uparrow$)} & \textbf{SSIM($\uparrow$)} & \textbf{NMSE($\downarrow$)} \\
\midrule
2  & $32.0\pm3.70$ & $0.890\pm0.048$ & $2.60\times10^{-2}\pm9.87\times10^{-3}$ \\
8  & $33.6\pm3.95$ & $0.900\pm0.048$ & $1.74\times10^{-2}\pm4.09\times10^{-3}$ \\
32 & $33.58\pm3.46$ & $0.900\pm0.041$ & $2.12\times10^{-2}\pm8.17\times10^{-3}$ \\
\bottomrule
\end{tabular}%
}
\end{table}

\section{Discussion}
\label{sec:discussion}

\subsection{Mantissa limits accuracy under MX}
The initial global power-of-two prescale and per-block MX scaling actively sets the dynamic range of each stage reducing chances of overflow/underflow errors. This suggests that quantization is the dominant error source from the butterfly complex multiplies. Thus formats with 3-bit mantissas (E4M3/E2M3) outperform 2-bit mantissas (E5M2/E3M2), despite the latter addressing a higher dynamic range. The round-trip experiment applies a second quantized transform, re-encodes the forward output, and introduces an additional $1/N^{2}$ normalization. Errors from the forward pass (e.g. rounding/clipping) are irrecoverable and get compounded during the inverse transform. Additionally per-block scales can differ between MXFFT and MXiFFT, adding extra requantization noise. This effect is seen visually and quantitatively in the increase in NMSE and drop in PSNR/SSIM relative to the forward-only experiment for the same formats.

\subsection{Block size trade-offs}
Small blocks ($B{=}2$; corresponding to a single complex value) produce many scale domains, which can result in frequent rescale/repack boundaries across stages amplifying rounding error. Blocks of $B{=}\{8,32\}$ amortize scale/metadata, stabilize amax, and can map better to vector/tensor hardware improving image quality and runtime speed. For the SKM-TEA dataset, using block sizes greater than $B{=}8$ results in more modest image quality gains.

\subsection{MRI evaluation}
SKM-TEA knee GRE exhibits coefficient distributions with more stable amax and fewer extremely small coefficients after prescale. Thus fewer terms land near FP8’s subnormal region under MX scaling. Conversely, fastMRI-brain FSE has a higher dynamic range resulting in heavier low-magnitude tails that can magnify rounding of small coefficients during the block scale. The inverse transform compounds this as two different block scales are calculated for the forward and inverse transform increasing quantization noise.

The results suggest that if the proposed mixed precision FFT implementation is to be incorporated to accelerate undersampled iterative image reconstruction \cite{Lustig2007, Aggarwal2019, https://doi.org/10.48550/arxiv.2004.06688}, where multiple FFT/iFFT pairs are invoked per iteration, mantissa width matters more than exponent range for retaining image quality. Preferred formats are MXFP8-E4M3 or MXFP6-E2M3 with $B{=}32$ and high-precision accumulation. Fusing encode$\rightarrow\mathcal{F}\rightarrow\mathcal{F}^{-1}\rightarrow$decode can further reduce round-trip loss by avoiding extra re-encodes.

\section{Conclusion}
\label{sec:conclusion}

%\section{Conclusion}
\vspace{-1mm}
We studied MX-scaled mixed-precision FFTs for MRI and quantified the impact on image fidelity across two public datasets. With a power-of-two global prescale and fixed per-block MX scaling, image quality depends primarily on mantissa precision rather than exponent range.  While the MXFP8-E4M3 format performs the best, these results are an intermediate step toward fast computation in more complex reconstruction pipelines and are not yet clinically ready for all domains. Our work suggests best practices are using FP32 accumulation with sufficiently large MX blocks for mixed precision MX-scaled FFTs.

\vfill\pagebreak

\newpage
% \section{REFERENCES}
% \label{sec:refs}

% List and number all bibliographical references at the end of the
% paper. The references can be numbered in alphabetic order or in
% order of appearance in the document. When referring to them in
% the text, type the corresponding reference number in square
% brackets as shown at the end of this sentence \cite{C2}. An
% additional final page (the fifth page, in most cases) is
% allowed, but must contain only references to the prior
% literature.

% Please follow the IEEE Citation Guidelines, \url{https://ieee-dataport.org/sites/default/files/analysis/27/IEEE\%20Citation\%20Guidelines.pdf} for formatting of references.

% % References should be produced using the bibtex program from suitable
% % BiBTeX files (here: strings, refs, manuals). The IEEEbib.bst bibliography
% % style file from IEEE produces unsorted bibliography list.
% % -------------------------------------------------------------------------
\bibliographystyle{IEEEbib}
\bibliography{strings,refs}

\end{document}